\documentclass[conference,twocolumn]{IEEEtran}

\usepackage{cite}
\usepackage[pdftex]{graphicx}
\usepackage{amsmath}

\usepackage{algorithm}
\makeatletter
\def\fs@ruledsafe{%
  \fs@ruled
  \def\@fs@pre{%
    \kern5pt
    \hrule height.8pt depth0pt
    \kern2pt
  }%
}
\floatstyle{ruledsafe}
\restylefloat{algorithm}
\makeatother
\usepackage{algpseudocode}
\algrenewcommand\algorithmicrequire{\textbf{Input:}}
\algrenewcommand\algorithmicensure{\textbf{Output:}}

\usepackage[caption=false,font=footnotesize]{subfig}
\usepackage{url}
\usepackage{amsthm}

\newtheoremstyle{mydefinition}
{}
{}
{}
{0pt}
{\bfseries}
{.}
{ }
{\thmname{#1}\thmnumber{ #2}: \thmnote{#3}}

\theoremstyle{mydefinition}

\newtheoremstyle{myremark}
{}
{}
{}
{0pt}
{\bfseries}
{.}
{ }
{\thmname{#1}\thmnumber{ #2}: \thmnote{#3}}
\theoremstyle{myremark}

\newtheoremstyle{remarkshort}
{}
{}
{}
{0pt}
{\bfseries}
{.}
{ }
{\thmname{#1}\thmnumber{ #2}}

\theoremstyle{remarkshort}

\theoremstyle{remarkshort}

\usepackage{amssymb,amsmath}
\usepackage{xspace}
\usepackage{bm}
\usepackage{bbm}
\usepackage{dsfont}
\usepackage{algorithm}
\usepackage{algpseudocode}

\let\originalleft\left
\let\originalright\right
\renewcommand{\left}{\mathopen{}\mathclose\bgroup\originalleft}
\renewcommand{\right}{\aftergroup\egroup\originalright}

\usepackage{xspace}
\usepackage[acronym,nogroupskip,nonumberlist,nopostdot]{glossaries}
\usepackage{relsize}
\usepackage{xcolor}

\newacronym{snr}{SNR}{signal-to-noise ratio}
\newacronym{sinr}{SINR}{signal-to-interference-plus-noise ratio}
\newacronym{inr}{INR}{interference-to-noise ratio}

\newcommand{\figref}[1]{Fig.~\ref{#1}}

\usepackage{mathtools}

\usepackage{enumitem}
\usepackage{graphicx}
\usepackage{tikz}
\usetikzlibrary{arrows.meta, positioning, calc, shapes.geometric}

\usepackage{pgfplots}
\pgfplotsset{compat=newest}
\usetikzlibrary{plotmarks}
\usetikzlibrary{arrows.meta}
\usepgfplotslibrary{patchplots}
\usepackage{grffile}

\pgfplotsset{plot coordinates/math parser=false}

\begin{document}
\raggedbottom
\title{Coexistence of LEO Satellite Communication Systems under Regulatory Protection Constraints}

\author{%
\IEEEauthorblockN{%
Sunwoo~Min\IEEEauthorrefmark{1}\IEEEauthorrefmark{2}, 
Daesik~Hong\IEEEauthorrefmark{2}, 
Jemin~Lee\IEEEauthorrefmark{2}, and 
Ian~P.~Roberts\IEEEauthorrefmark{1}}%
\IEEEauthorblockA{\IEEEauthorrefmark{1}Department of Electrical and Computer Engineering, UCLA, Los Angeles, CA USA}%
\IEEEauthorblockA{\IEEEauthorrefmark{2}School of Electrical and Electronic Engineering, Yonsei University, Seoul, South Korea}%
}

\maketitle

\begin{abstract}
This paper investigates the coexistence of two low Earth orbit (LEO) satellite communication systems sharing the same frequency band: a \textit{primary} system with regulatory precedence over the spectrum and a \textit{secondary} system that must not cause excessive interference to the primary system's ground users.
This spectrum sharing paradigm was recently formalized by the United States Federal Communications Commission (FCC) through short-term and long-term \textit{protection criteria} that limit primary-link unavailability and throughput degradation due to interference inflicted by the secondary system.
To facilitate coexistence, we propose a satellite selection mechanism that associates the secondary system's satellites with clusters of ground cells at each handover to maximize secondary-user throughput while abiding by these regulatory constraints.
Our mechanism enforces both constraints through a cumulative assessment that combines past primary-link performance with predicted performance over the upcoming handover interval.
Simulation results show that our proposed method achieves up to 31\% higher secondary-user throughput than prior work that imposes \gls{inr} constraints, even when its constraints are tuned to satisfy the FCC's throughput-based protection criteria.
\end{abstract}


\glsresetall

\section{Introduction}
Low Earth orbit (LEO) satellite systems deployed by operators such as SpaceX, Amazon, and OneWeb share portions of the Ka-band spectrum under non-exclusive spectrum allocations by the United States Federal Communications Commission (FCC)~\cite{roberts_2026_spectrum}.
The FCC groups applications for satellite system authorization into \textit{processing rounds} based on their filing dates~\cite{FCC24117,roberts_2026_spectrum}.
Under this framework, the FCC grants spectrum priority to systems authorized in earlier processing rounds, referred to as \textit{primary} systems, over those authorized in later rounds, referred to as \textit{secondary} systems.
In the absence of a coordination agreement, the secondary system must protect primary ground users from excessive interference.

To mitigate interference between coexisting LEO satellite systems, prior work has proposed adaptive receive beamforming/nulling at ground users~\cite{attbf,smartut}.
While these methods can be effective, they naturally face limitations when a user's serving satellite is in direct angular alignment with an interfering satellite.
Satellite selection complements these receive beam control techniques by exploiting the spatial diversity of available satellites to increase the angular separation between serving and interfering satellites as viewed from ground users, thereby reducing interference.
This concept was investigated in~\cite{kim2025feasibility}, which examined the feasibility of in-band coexistence through secondary satellite selection and showed that multiple secondary satellites can often satisfy \gls{inr} constraints for primary-link protection.
Recently, the work in~\cite{roberts_selection} proposed a secondary satellite selection mechanism to enhance secondary-user throughput while protecting primary ground users under time-averaged and instantaneous \gls{inr} constraints.

To formalize inter-round protection without a coordination agreement, the FCC recently adopted throughput-based protection criteria in late 2024~\cite{FCC24117}.
The short-term criterion limits the increase in primary-link unavailability to 0.4\%, while the long-term criterion limits the average throughput degradation of primary links to 3\%~\cite{FCC24117}.
Unlike \gls{inr}-based criteria, these criteria directly limit the degradation in primary-link performance caused by secondary transmissions.
An \gls{inr} value alone does not determine whether these criteria are satisfied, since the resulting performance degradation also depends on the primary-link \gls{snr}.

In this work, we propose a satellite selection mechanism for a secondary satellite operator that explicitly considers the FCC's newly adopted throughput-based protection constraints.
Given the primary satellite associations, our proposed mechanism updates secondary satellite-to-cluster associations at handover instants to maximize secondary-user throughput while satisfying these constraints.
To enforce both short-term and long-term criteria, we employ a cumulative assessment that combines past primary-link performance with predicted performance over the upcoming handover interval under each candidate secondary association.

\section{System Model}
\label{sec:system_model}

\subsection{Network Geometry and Satellite-to-Cluster Associations}
We consider two dense LEO satellite communication systems, the primary and secondary systems, that share the same frequency band and independently provide downlink service to their respective ground users within their overlapping coverage area.
Let $\mathcal{P}^{t}=\{p_i^{t}:i=1,\ldots,N_\mathrm{P}\}$ and $\mathcal{S}^{t}=\{s_k^{t}:k=1,\ldots,N_\mathrm{S}\}$ denote the sets of all satellites in the primary and secondary constellations at time $t$, respectively.

\begin{figure}[!t]
    \centering
    \includegraphics[width=\linewidth]{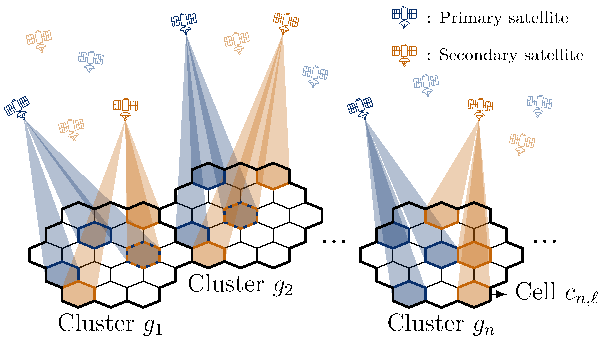}
    \caption{Satellite-to-cluster associations and spot-beam transmissions in coexisting primary and secondary LEO satellite communication systems.}
    \label{fig:assoc}
\end{figure}

As shown in \figref{fig:assoc}, the geographical area is partitioned into a set $\mathcal{G}=\{g_n:n=1,\ldots,N_\mathrm{G}\}$ of $N_\mathrm{G}$ clusters, with each cluster $g_n=\{c_{n,\ell}:\ell=1,\ldots,N_\mathrm{C}\}$ containing $N_\mathrm{C}$ cells.
In practical dense LEO satellite communication systems, a ground cell may have multiple candidate serving satellites, resulting in many possible satellite-to-cell associations across the network.
As in~\cite{roberts_selection}, we consider satellite associations at the cluster level rather than the individual cell level.
At time $t$, the primary and secondary candidate satellite sets for cluster $g_n$ are defined using the minimum elevation angle required for communication, $\epsilon_{\min}$, as
\begin{align}
\mathcal{P}_n^{t}
&=\left\{p_i^t\in\mathcal{P}^{t}:
\epsilon(g_n;p_i^t)\geq\epsilon_{\min}\right\},
\label{eq:primary_candidate_set}\\
\mathcal{S}_n^{t}
&=\left\{s_k^t\in\mathcal{S}^{t}:
\epsilon(g_n;s_k^t)\geq\epsilon_{\min}\right\},
\label{eq:secondary_candidate_set}
\end{align}
where $\epsilon(g_n;z)$ denotes the minimum elevation angle of satellite $z$ over all cell centers in cluster $g_n$.
Given the primary satellite associations, we investigate how to select secondary satellites from the secondary candidate satellite sets and associate them with clusters while protecting primary ground users.
At each handover instant $\tau$, the secondary satellite-to-cluster associations are updated and remain fixed over the handover interval $\mathcal{T}^{\tau}=[\tau,\tau+T_\mathrm{H})$, where $T_\mathrm{H}$ denotes the handover period.
The corresponding secondary satellite association matrix is defined as $\mathbf{X}^{\tau}=[x_{k,n}^{\tau}]\in\{0,1\}^{N_\mathrm{S}\times N_\mathrm{G}}$, whose $(k,n)$-th entry $x_{k,n}^{\tau}$ equals $1$ if secondary satellite $s_k^{\tau}$ is associated with cluster $g_n$ and $0$ otherwise.
Thus, the association variable at time $t$ satisfies $x_{k,n}^{t}=x_{k,n}^{\tau}$ for all $t\in\mathcal{T}^{\tau}$.
Within each system, we assume that each satellite is associated with at most one cluster and each cluster is associated with at most one satellite at a time.

Because a cluster contains $N_\mathrm{C}$ cells, each satellite serves its associated cluster via beam hopping, illuminating up to $N_\mathrm{B}$ cells at a time using highly directional spot beams directed toward the selected cell centers.
Accordingly, let $y_{k,n,\ell}^{t}\in\{0,1\}$ indicate whether secondary satellite $s_k^t$ illuminates cell $c_{n,\ell}$ at time $t$ when associated with cluster $g_n$, where $\sum_{\ell=1}^{N_\mathrm{C}}y_{k,n,\ell}^{t}\leq N_\mathrm{B}$.
We assume that beam-hopping schedules are coordinated within each system to mitigate intra-system interference, and we neglect the residual interference in the \gls{inr} and \gls{sinr} calculations.

\subsection{Link-Quality Metrics}
Given the satellite-to-cluster associations and beam-hopping schedules, we characterize the desired signal and aggregate interference from the other system at each user.
For both systems, we assume line-of-sight propagation between satellites and users.
Doppler effects due to the rapid motion of LEO satellites are assumed to be compensated, with negligible residual impact on link performance.

Let $L(u,z)$ denote the path loss from satellite $z$ to user $u$.
Let $G_{\mathrm{tx}}(u,z;c)$ denote the transmit antenna gain of satellite $z$ toward user $u$ when its beam is steered toward the center of cell $c$, and let $G_{\mathrm{rx}}(u,p_i^t)$ denote the receive antenna gain of user $u$ toward its serving satellite $p_i^t$.
Let $P_{\mathrm{tx}}(z)$ denote the transmit power per active beam of satellite $z$, and let $P_{\mathrm{n}}(u)$ denote the receiver noise power at user $u$.
Let $c_u$ and $c_v$ denote the cells containing primary ground user $u$ and secondary ground user $v$, respectively.
For a primary ground user $u$ served by satellite $p_i^t$, the \gls{snr} at time $t$ is given by
\begin{align}
\mathsf{SNR}^{t}(u,p_i^t,c_u)
&=
\frac{
P_{\mathrm{tx}}(p_i^t)
G_{\mathrm{tx}}(u,p_i^t;c_u)
G_{\mathrm{rx}}(u,p_i^t)
}{
P_{\mathrm{n}}(u)L(u,p_i^t)
}.
\label{eq:primary_snr}
\end{align}

For an interfering secondary satellite $s_k^t$, let $G_{\mathrm{rx}}(u,p_i^t;s_k^t)$ denote the receive antenna gain toward satellite $s_k^t$ when the receive beam of user $u$ is steered toward its serving satellite $p_i^t$.
Using the satellite association indicators $x_{k,n}^t$ and cell illumination indicators $y_{k,n,\ell}^t$, the aggregate \gls{inr} at primary ground user $u$ served by satellite $p_i^t$ at time $t$ can be expressed as
\begin{align}
\mathsf{INR}^{t}(u,p_i^t)
&\triangleq
\sum_{k=1}^{N_\mathrm{S}}
\sum_{n=1}^{N_\mathrm{G}}
\sum_{\ell=1}^{N_\mathrm{C}}
x^{t}_{k,n}y^{t}_{k,n,\ell}
\notag\\
&\quad\cdot
\frac{
P_{\mathrm{tx}}(s_k^t)
G_{\mathrm{tx}}(u,s_k^t;c_{n,\ell})
G_{\mathrm{rx}}(u,p_i^t;s_k^t)
}{
P_{\mathrm{n}}(u)L(u,s_k^t)
}.
\label{eq:primary_inr}
\end{align}

The corresponding \gls{sinr} at time $t$ is given by
\begin{align}
\mathsf{SINR}^{t}(u,p_i^t,c_u)
&=
\frac{
\mathsf{SNR}^{t}(u,p_i^t,c_u)
}{
1+\mathsf{INR}^{t}(u,p_i^t)
}.
\label{eq:primary_sinr}
\end{align}
For the secondary system, the \gls{snr} and \gls{inr} are defined analogously.
The \gls{sinr} of secondary ground user $v$ served by satellite $s_k^t$ is given by
\begin{align}
\mathsf{SINR}^{t}(v,s_k^t,c_v)
&=
\frac{
\mathsf{SNR}^{t}(v,s_k^t,c_v)
}{
1+\mathsf{INR}^{t}(v,s_k^t)
}.
\label{eq:secondary_SINR}
\end{align}

\section{Proposed Secondary Satellite Selection}

\subsection{Primary Throughput-Based Protection Criteria}
In 2024, the FCC adopted short-term and long-term protection criteria under 47 CFR § 25.261(d)~\cite{FCC24117}, limiting increases in link unavailability and time-weighted average throughput degradation, respectively.
For a given evaluation interval $T$, the two criteria are defined as follows.\footnote{%
The protection criteria apply to primary links whose baseline availability is at least $99.0\%$ at a $0$~dB carrier-to-noise ratio (C/N) threshold, which all primary links considered in this paper satisfy.
Throughout this paper, \gls{snr} denotes the C/N at the ground user receiver.
}
For brevity, let $\mathsf{SNR}_u(t)\triangleq\mathsf{SNR}^t(u,p_i^t,c_u)$ and $\mathsf{SINR}_u(t)\triangleq\mathsf{SINR}^t(u,p_i^t,c_u)$ denote the \gls{snr} and \gls{sinr} of primary ground user $u$ at time~$t$.

\begin{itemize}
\item \textbf{Absolute Increase in Link Unavailability:}
Link unavailability is the fraction of time during which a primary link's \gls{snr} (without secondary interference) or \gls{sinr} (with secondary interference) falls below the outage threshold (assumed to be $0$~dB per the FCC~\cite{FCC24117}).
For each primary link, the absolute increase in link unavailability due to secondary transmissions must not exceed $0.4\%$.
With $\mathds{1}\{\cdot\}$ denoting the indicator function, $|T|$ denoting the duration of the evaluation interval, and $0$~dB corresponding to $1$ on a linear scale, this criterion is expressed as
\begin{align}
\delta_{\mathrm{L},u}
&\triangleq
\frac{1}{|T|}
\int_T
\Big[
\mathds{1}\left\{\mathsf{SINR}_u(t)<1\right\}
\notag\\
&\qquad -
\mathds{1}\left\{\mathsf{SNR}_u(t)<1\right\}
\Big]
\,\mathrm{d}t
\leq 0.004.
\label{eq:unavailability_constraint}
\end{align}

\item \textbf{Time-Weighted Average Throughput Degradation:}
The relative reduction in the time-weighted average throughput of each primary link due to secondary transmissions must not exceed $3\%$.
The function $R(\cdot)$ maps instantaneous link quality to throughput.\footnote{For both systems, we model throughput as $R(\gamma)=B\log_2(1+\gamma)$, where $B$ is the bandwidth and $\gamma$ is the \gls{snr} or \gls{sinr}.}
This criterion is expressed as
\begin{align}
\delta_{\mathrm{T},u}
&\triangleq
1-
\frac{
\int_T R\left(\mathsf{SINR}_u(t)\right)\,\mathrm{d}t
}{
\int_T R\left(\mathsf{SNR}_u(t)\right)\,\mathrm{d}t
}
\leq 0.03.
\label{eq:throughput_degradation_constraint}
\end{align}
\end{itemize}

\begin{figure}[!t]
    \centering
    \includegraphics[width=\linewidth]{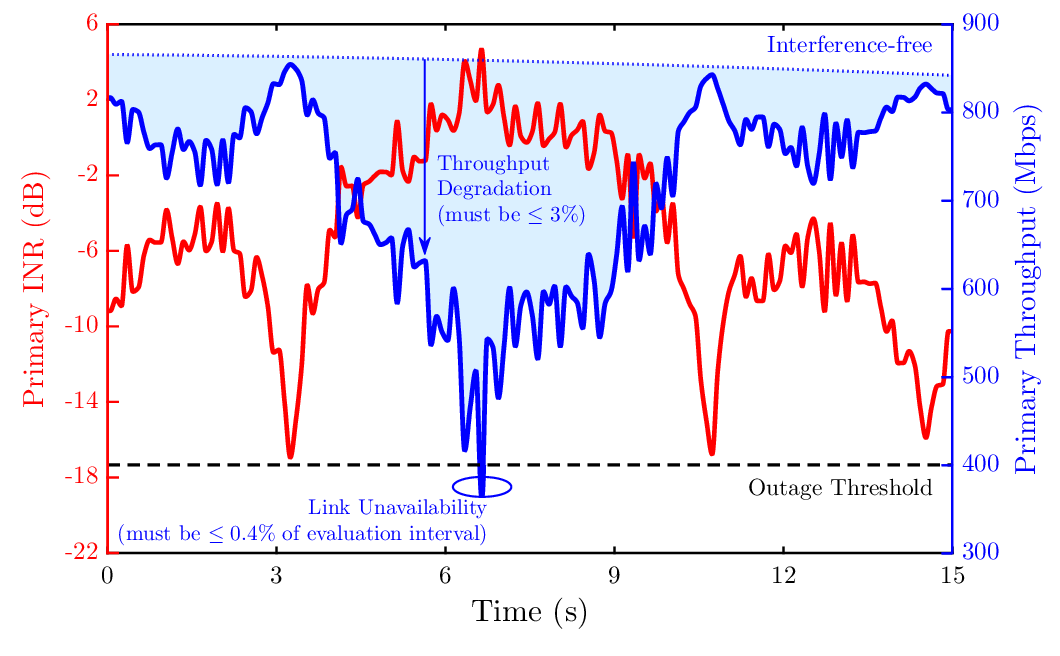}
    \caption{\gls{inr} and throughput at a primary ground user over a $15$\,s handover interval under fixed satellite associations determined by highest-elevation primary satellite selection and random secondary satellite selection, with $N_\mathrm{G}=20$ and $N_\mathrm{B}=16$.}
    \label{fig:primary_inr}
\end{figure}

\figref{fig:primary_inr} shows the \gls{inr} and corresponding throughput of a primary ground user under highest-elevation primary satellite selection and random secondary satellite selection without primary protection constraints.
The black dashed line marks the outage threshold, while the light-blue shaded area illustrates the throughput loss relative to the interference-free case.
In this particular realization, interference increases link unavailability from $0\%$ to $3.2\%$ and causes an average throughput degradation of $16.7\%$, exceeding the respective protection limits of $0.4\%$ and $3\%$.
These violations highlight the need for the proposed method, which explicitly enforces both primary-link performance constraints when determining secondary satellite associations.

\subsection{Secondary Satellite Selection}
For simplicity, we consider synchronized handovers across both systems.
At each handover instant $\tau$, the secondary operator optimizes its satellite-to-cluster associations $\mathbf X^\tau$ given the primary satellite associations over $\mathcal T^\tau$.
Using known satellite trajectories and antenna patterns, we assume that the secondary system can reliably predict the \glspl{snr} and \glspl{sinr} of primary and secondary ground users via \eqref{eq:primary_snr}--\eqref{eq:secondary_SINR} under each potential $\mathbf X^\tau$ over $\mathcal T^\tau$.

Let $\mathcal C=\bigcup_{n=1}^{N_\mathrm{G}}g_n$ denote the set of ground cells, and let $u_c$ denote the representative primary ground user located at the center of cell $c\in\mathcal C$.
For primary-link protection, we evaluate the \gls{snr} and \gls{sinr} at each $u_c$, assuming that its serving primary satellite directs a spot beam toward the center of cell $c$.
Using the resulting \glspl{snr} and \glspl{sinr}, we evaluate the FCC protection criteria in \eqref{eq:unavailability_constraint} and \eqref{eq:throughput_degradation_constraint} for each representative primary link over the cumulative evaluation interval $[0,\tau+T_\mathrm{H})$.
The primary-link performance accumulated before $\tau$ is already determined, while the predicted performance over the upcoming handover interval $\mathcal T^\tau$ depends on $\mathbf X^\tau$.
We denote the resulting increase in link unavailability and throughput degradation by $\delta_{\mathrm{L},u_c}(\mathbf X^\tau)$ and $\delta_{\mathrm{T},u_c}(\mathbf X^\tau)$, respectively.

For each handover interval $\mathcal{T}^{\tau}$, the time-average aggregate secondary-user throughput when satellite $s_k$ serves cluster $g_n$ is defined as
\begin{align}
\overline R_{k,n}^{\tau}
&=\frac{1}{T_\mathrm{H}}\int_{\mathcal{T}^{\tau}}\!\sum_{v\in\mathcal{V}_{k,n}^{t,\mathrm{act}}}\!R\bigl(\mathsf{SINR}^{t}(v,s_k^t,c_v)\bigr)\,\mathrm{d}t,
\label{eq:secondary_throughput}
\end{align}
where $\mathcal{V}_{k,n}^{t,\mathrm{act}}$ denotes the set of secondary ground users that would be served at time $t$ using up to $N_\mathrm{B}$ active spot beams if satellite $s_k$ were assigned to cluster $g_n$.

Our goal is to find the secondary satellite association that maximizes the average aggregate secondary-user throughput over $\mathcal T^\tau$ while satisfying the FCC protection constraints and the satellite-to-cluster association constraints.
Accordingly, the optimization problem at handover instant $\tau$ is formulated as follows:
\begingroup
\setlength{\jot}{5pt}
\begin{subequations}\label{eq:optimization}
\begin{align}
\underset{\mathbf X^\tau}{\mathrm{maximize}}\quad
& \sum_{k=1}^{N_\mathrm{S}}
  \sum_{n=1}^{N_\mathrm{G}}
  \overline R_{k,n}^\tau x_{k,n}^\tau
\label{eq:optimization_a}\\
\mathrm{s.t.}\quad
& \delta_{\mathrm{L},u_c}(\mathbf X^\tau)\leq 0.004,
  && \forall c\in\mathcal C,
\label{eq:optimization_b}\\
& \delta_{\mathrm{T},u_c}(\mathbf X^\tau)\leq 0.03,
  && \forall c\in\mathcal C,
\label{eq:optimization_c}\\
& \sum_{n=1}^{N_\mathrm{G}}x_{k,n}^\tau\leq 1,
  && \forall k,
\label{eq:optimization_d}\\
& \sum_{k=1}^{N_\mathrm{S}}x_{k,n}^\tau\leq 1,
  && \forall n,
\label{eq:optimization_e}\\
& x_{k,n}^\tau\in\{0,1\},
  && \forall k,\forall n.
\label{eq:optimization_f}
\end{align}
\end{subequations}
\endgroup
Constraints \eqref{eq:optimization_b} and \eqref{eq:optimization_c} enforce the primary-link protection criteria on the absolute increase in link unavailability and throughput degradation, respectively.
Constraints \eqref{eq:optimization_d} and \eqref{eq:optimization_e} allow at most one association per satellite and cluster, respectively, while \eqref{eq:optimization_f} restricts the association variables to binary values, with $x_{k,n}^\tau=0$ if $s_k^t\notin\mathcal S_n^t$ for any $t\in\mathcal T^\tau$.
Algorithm~\ref{alg:proposed} summarizes the proposed secondary satellite selection procedure at each handover instant $\tau$.

\begin{algorithm}[t]
\caption{Throughput-Based Secondary Satellite Selection.}
\label{alg:proposed}
\begin{algorithmic}[1]
\Statex \textbf{At handover instant $\tau$:}
\Require Primary satellite associations, satellite trajectories, and accumulated primary-link performance
\Ensure Secondary satellite association matrix $\widehat{\mathbf X}^{\tau}$

\Statex \textbf{Step 1: Predict primary-link and secondary-link performance}
\State For each cluster $g_n$, determine candidate secondary satellites satisfying the minimum elevation angle requirement at each $t\in\mathcal T^\tau$ using \eqref{eq:secondary_candidate_set}, i.e., $\mathcal{S}_n^{t}$.
\State Compute $\overline R_{k,n}^\tau$ for each $g_n$ and candidate secondary satellite $s_k$ using \eqref{eq:secondary_throughput}.
\State Compute primary-link \glspl{sinr} for each potential secondary satellite association matrix $\mathbf X^{\tau}$, for all $c\in\mathcal C$ over $\mathcal T^\tau$ using \eqref{eq:primary_snr}--\eqref{eq:primary_sinr}.

\Statex \textbf{Step 2: Optimize secondary satellite-to-cluster associations}
\State Form $\delta_{\mathrm{L},u_c}(\mathbf X^\tau)$ and $\delta_{\mathrm{T},u_c}(\mathbf X^\tau)$ for each potential $\mathbf X^\tau$ using accumulated and predicted primary-link performance for all $c\in\mathcal C$.
\State Solve \eqref{eq:optimization} using SCIP via PySCIPOpt~\cite{pyscipopt} to obtain $\widehat{\mathbf X}^{\tau}$.

\Statex \textbf{Step 3: Apply associations and update primary-link and secondary-link performance}
\State Apply $\widehat{\mathbf X}^{\tau}$ over $\mathcal T^\tau$.
\State Update accumulated primary-link performance for the next handover.
\State \Return $\widehat{\mathbf X}^{\tau}$.
\end{algorithmic}
\end{algorithm}

\begin{figure*}[!t]
    \centering
    \subfloat[CDF of primary SINR.]{%
        \includegraphics[width=0.49\textwidth]{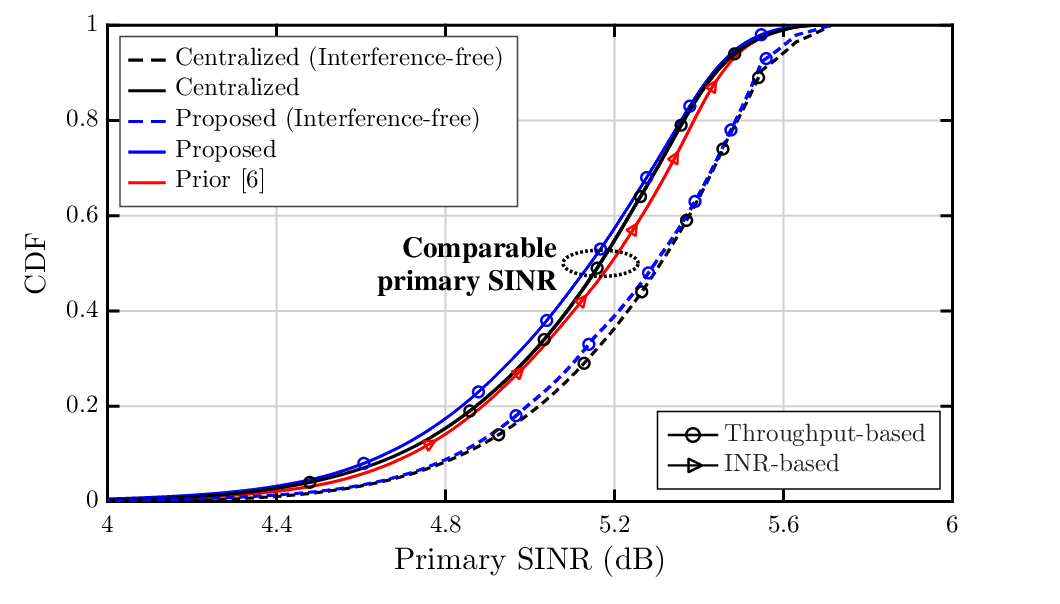}
        \label{fig:pri_cdf_final}
    }
    \hfill
    \subfloat[CDF of secondary SINR.]{%
        \includegraphics[width=0.49\textwidth]{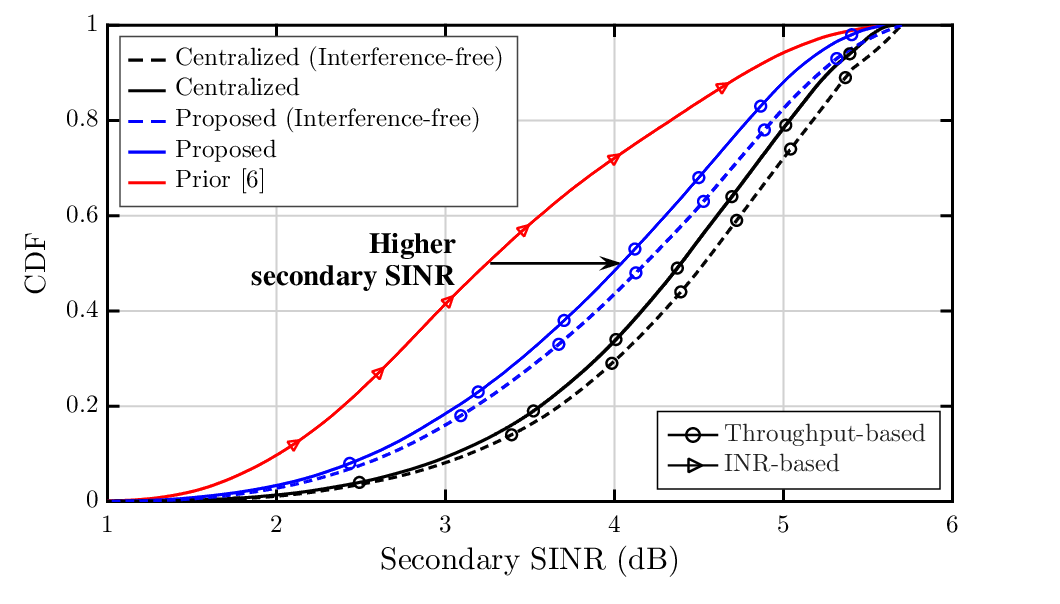}
        \label{fig:sec_cdf_final}
    }
    \caption{CDFs of primary and secondary SINRs across actively served ground users and time under different satellite selection methods, with $N_\mathrm{G}=20$ clusters and $N_\mathrm{B}=16$ active spot beams per serving satellite.}
    \label{fig:sinr_cdf}
\end{figure*}
\begin{figure}[!t]
    \centering
    \includegraphics[width=\linewidth]{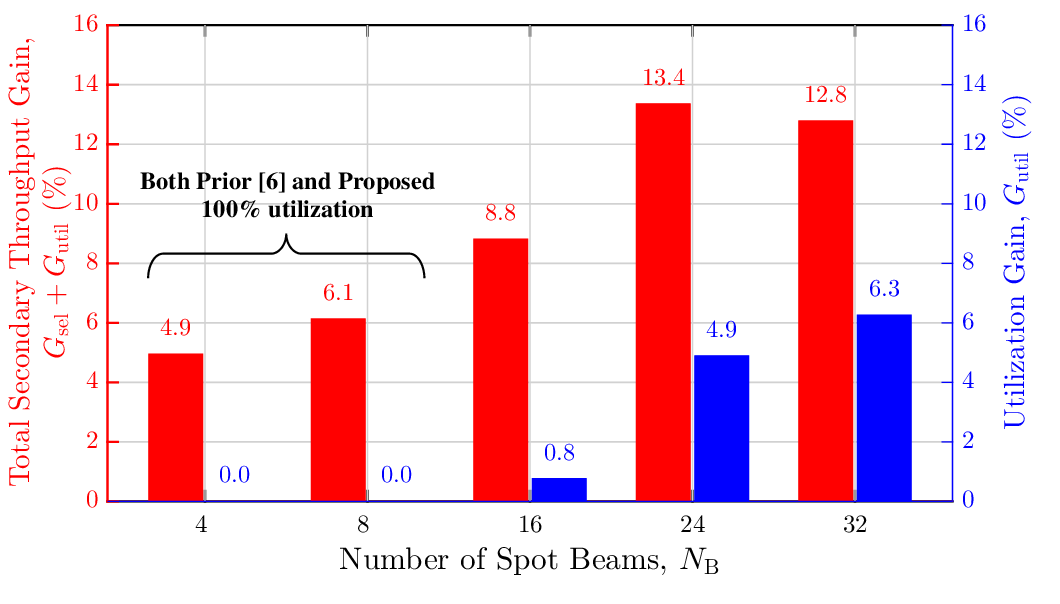}
    \caption{Total secondary throughput gain, $G_{\mathrm{sel}}+G_{\mathrm{util}}$, of the proposed method over the prior method and its utilization component, $G_{\mathrm{util}}$, versus the maximum number of active spot beams per satellite, with $N_\mathrm{G}=20$.
    As $N_{\mathrm B}$ varies, each satellite's total transmit power is fixed at its default-configuration value and equally divided among its active spot beams.}
    \label{fig:gain_final}
\end{figure}

\section{Performance Evaluation}
\subsection{Simulation Setup and Baselines}
To assess our proposed scheme, we consider a hypothetical inter-round coexistence scenario using FCC-filed constellation geometries of Starlink Gen1 and Gen2~\cite{star_constell1,star_constell2,star_constell_gen2} as the primary system ($N_\mathrm{P}=11{,}908$) and Kuiper~\cite{kuip_constell1} as the secondary system ($N_\mathrm{S}=3{,}236$).
Unless otherwise specified, the region contains $N_\mathrm{G}=20$ clusters, each with $N_\mathrm{C}=127$ Earth-fixed hexagonal cells of approximately $10$\,km edge length~\cite{earthfixed,h3}.
Each cell has one representative user from each system, both located at its center.
Candidate satellites are defined as those with elevation angles of at least $\epsilon_{\min}=35^\circ$ at every cell center in a cluster~\cite{kuip_constell1}.
Each serving satellite illuminates $N_\mathrm{B}=16$ randomly selected cells in its associated cluster.
Satellite positions are sampled every $0.02$\,s over $T=120$\,s, with associations updated every $T_\mathrm{H}=15$\,s~\cite{handover}.

We assume free-space path loss and neglect atmospheric attenuation and scintillation~\cite{kim2025feasibility,roberts_selection}.
Both systems share $400$\,MHz at $20$\,GHz and employ $64\times64$ satellite and $32\times32$ user uniform planar arrays with half-wavelength spacing.
User antenna panels face the zenith, with receive beams steered toward serving satellites.
For the default configuration, the maximum per-beam EIRP spectral densities are $-54.3$\,dBW/Hz~\cite{starlink_eirp} and $-53.3$\,dBW/Hz~\cite{kuiper_eirp} for the primary and secondary systems, respectively.
The receiver noise spectral density is $-174$\,dBm/Hz, with a noise figure of $1.2$\,dB~\cite{3gpp_tr38811}.

Under our proposed method, the primary system associates each cluster with its highest-elevation available satellite at each handover while the secondary system optimizes its satellite associations via Algorithm~\ref{alg:proposed}.
We compare our proposed method against the following two benchmarks:
\begin{itemize}
    \item Prior~\cite{roberts_selection}:
    Uses the same highest-elevation primary satellite associations, but selects secondary satellites in each handover interval under time-averaged and instantaneous \gls{inr} thresholds tuned to maximize secondary throughput while satisfying the FCC protection criteria.
    \item Centralized:
    Jointly optimizes both primary and secondary satellite associations to maximize their sum throughput under the same FCC protection criteria, representing the best-case performance under cross-system coordination.
\end{itemize}

\subsection{Components of Secondary Throughput Gain}
We decompose the secondary throughput gain into satellite-selection and utilization gains.
Utilization is defined as the percentage of clusters associated with a secondary satellite, averaged over all handover intervals.
For each method $m\in\{\mathrm{Prior},\mathrm{Prop}\}$, let $\overline{R}^m$ denote the average aggregate secondary throughput over the total simulation period.
Let $\overline{R}_{\mathrm{com}}^m$ and $\overline{R}_{\mathrm{only}}^m$ denote the corresponding contributions from cluster--interval pairs served by both methods and only by method $m$, respectively.
Thus, $\overline{R}^m=\overline{R}_{\mathrm{com}}^m+\overline{R}_{\mathrm{only}}^m$.
Both gain components are normalized by the prior method's average aggregate secondary throughput, $\overline{R}^{\mathrm{Prior}}$,
and are defined as follows:
\begin{enumerate}
    \item \textbf{Satellite-selection gain}:
    The normalized throughput difference over all cluster--interval
    pairs served by both methods is given by
    \begin{equation}
        G_{\mathrm{sel}}
        =
        \frac{
        \overline{R}_{\mathrm{com}}^{\mathrm{Prop}}
        -
        \overline{R}_{\mathrm{com}}^{\mathrm{Prior}}
        }{
        \overline{R}^{\mathrm{Prior}}
        }.
        \label{eq:satellite_selection_gain}
    \end{equation}

    \item \textbf{Utilization gain}:
    The normalized throughput difference over all cluster--interval
    pairs served by exactly one of the two methods is given by
    \begin{equation}
        G_{\mathrm{util}}
        =
        \frac{
        \overline{R}_{\mathrm{only}}^{\mathrm{Prop}}
        -
        \overline{R}_{\mathrm{only}}^{\mathrm{Prior}}
        }{
        \overline{R}^{\mathrm{Prior}}
        }.
        \label{eq:association_utilization_gain}
    \end{equation}
\end{enumerate}
The total secondary throughput gain is then
\begin{equation}
\frac{
\overline{R}^{\mathrm{Prop}}
-
\overline{R}^{\mathrm{Prior}}
}{
\overline{R}^{\mathrm{Prior}}
}
=
G_{\mathrm{sel}}+G_{\mathrm{util}}.
\label{eq:throughput_gain_decomposition}
\end{equation}

\subsection{Simulation Results}
\figref{fig:sinr_cdf} compares the primary and secondary \gls{sinr} distributions across actively served ground users under different satellite selection methods.
The interference-free curves for the centralized and proposed methods are obtained by removing interference while retaining their respective satellite-to-cluster associations.
Under the same primary satellite associations and FCC protection criteria, the proposed method achieves primary \glspl{sinr} comparable to those of the prior method, as shown in \figref{fig:sinr_cdf}(a), while achieving higher secondary \glspl{sinr}, as shown in \figref{fig:sinr_cdf}(b).
The proposed method also yields a secondary \gls{sinr} distribution close to that of the centralized method, despite optimizing only secondary satellite-to-cluster associations.
Since these CDFs include only actively served users, they do not capture the utilization gain $G_{\mathrm{util}}$.
We next examine how the satellite-selection gain $G_{\mathrm{sel}}$ and utilization gain $G_{\mathrm{util}}$ contribute to the total secondary throughput gain.

\figref{fig:gain_final} shows the total secondary throughput gain of the proposed method over the prior method~\cite{roberts_selection} and the contribution from utilization for different numbers of spot beams.
When $N_\mathrm{B}$ is small, both methods can associate every cluster with a secondary satellite in each handover interval while satisfying their respective protection constraints, achieving 100\% utilization.
At $N_\mathrm{B}=4$ and $8$, $G_{\mathrm{util}}$ is therefore zero, and the $4.9\%$ and $6.1\%$ throughput improvements arise solely from the satellite-selection gain $G_{\mathrm{sel}}$.
As $N_\mathrm{B}$ increases under a fixed total transmit power per satellite, the per-beam transmit power decreases, reducing the primary-link \gls{snr} and its margin for tolerating interference.
To satisfy the FCC protection criteria, the prior method's selected \gls{inr} thresholds become more stringent as $N_\mathrm{B}$ increases, restricting feasible satellite associations and reducing utilization.
However, the proposed method directly enforces the primary-link performance constraints, allowing more feasible satellite-to-cluster associations.
Consequently, $G_{\mathrm{util}}$ grows from $0.8\%$ at $N_\mathrm{B}=16$ to $6.3\%$ at $N_\mathrm{B}=32$, yielding total throughput gains of up to $13.4\%$ (at $N_\mathrm{B}=24$).
\begin{figure}[!t]
    \centering
    \includegraphics[width=\linewidth]{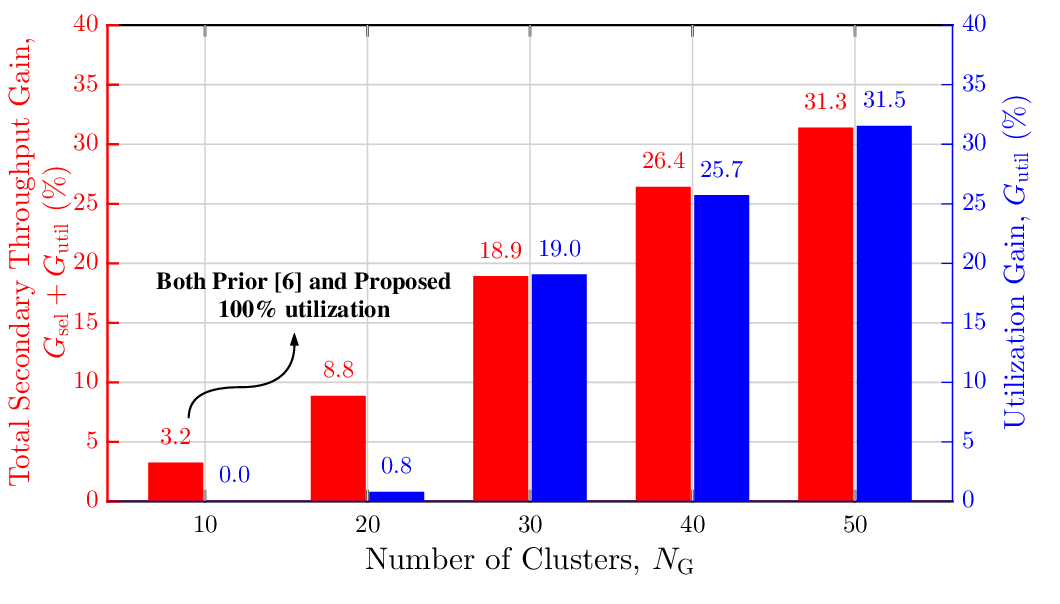}
    \caption{Total secondary throughput gain, $G_{\mathrm{sel}}+G_{\mathrm{util}}$, of the proposed method over the prior method and its utilization component, $G_{\mathrm{util}}$, versus the number of clusters, with $N_\mathrm{B}=16$.}
    \label{fig:clus}
\end{figure}

\figref{fig:clus} shows the total secondary throughput gain of the proposed method over the prior method and the contribution from utilization, $G_{\mathrm{util}}$, for different numbers of clusters.
As the number of clusters increases, utilization decreases for both methods, but it decreases more rapidly for the prior method, widening the utilization gap between the two methods.
The proposed method directly enforces the FCC protection criteria and exploits protection margins accumulated over previous intervals.
This enables additional secondary satellite-to-cluster associations that satisfy these cumulative limits but are excluded by the prior method's \gls{inr}-based constraints.
Consequently, the utilization gain $G_{\mathrm{util}}$ becomes the dominant contributor to the total secondary throughput gain as the number of clusters increases.
At $N_\mathrm{G}=30$ and $50$, $G_{\mathrm{util}}$ slightly exceeds the total gain ($G_{\mathrm{sel}}<0$), as serving additional clusters leaves fewer candidate satellites and less primary protection margin for clusters served by both methods, though this minor trade-off yields up to a $31.3\%$ total throughput gain.

\section{Conclusion}
\label{sec:conclusion}
This paper presented a secondary satellite selection mechanism for the in-band coexistence of primary and secondary LEO satellite communication systems under the FCC's recently adopted throughput-based protection criteria.
To limit both primary-link unavailability and average throughput degradation, our mechanism employs a cumulative assessment at each handover that combines past primary-link performance with predicted performance over the upcoming handover interval.
Simulation results demonstrated that directly enforcing these performance criteria allows the secondary system to select more favorable satellites and serve more ground clusters than an \gls{inr}-based benchmark tuned to meet the same FCC limits, improving secondary-user throughput by up to 31.3\%.
Future work could extend this framework to coexistence among multiple secondary constellations and relax the assumption of known primary satellite associations.

\bibliographystyle{bibtex/IEEEtran}
{\small \bibliography{bibtex/IEEEabrv,refs}}

@STRING{IEEE_J_WCOM       = "{IEEE} Trans. Wireless Commun."}

@article{roberts_selection,
  author  = {Eunsun Kim and Ian P. Roberts and Taekyun Lee and Jeffrey G. Andrews},
  title   = {Satellite Selection for In-Band Coexistence of Dense {LEO} Networks},
  journal = IEEE_J_WCOM,
  year    = {2026},
  volume  = {25},
  pages   = {10274--10289},
  doi     = {10.1109/TWC.2025.3645992}
}

@techreport{3gpp_tr38811,
  author      = {{3GPP}},
  title       = {Study on New Radio ({NR}) to Support Non-Terrestrial Networks ({NTN})},
  type        = {Tech. Rep.},
  number      = {38.811},
  year        = {2020},
  month       = jul
}

@misc{roberts_2026_spectrum,
  author       = {I. P. Roberts},
  title        = {Spectrum Sharing Among Broadband {LEO} Mega-Constellations: Interference Regulations and Coexistence Mechanisms},
  howpublished = {arXiv preprint arXiv:2609.36122},
  year         = {2026},
  month        = sep
}

@misc{FCC24117,
  author       = {{Federal Communications Commission}},
  title        = {Revising Spectrum Sharing Rules for Non-Geostationary Orbit, Fixed-Satellite Service Systems},
  howpublished = {FCC 24-117},
  year         = {2024},
  month        = nov
}

@inproceedings{handover,
  author    = {N. Mohan and others},
  title     = {A Multifaceted Look at {Starlink} Performance},
  booktitle = {Proc. ACM Web Conf.},
  year      = {2024},
  month     = may,
  pages     = {2723--2734}
}

@article{attbf,
  author  = {Almoatssimbillah Saifaldawla and Flor Ortiz and
             Eva Lagunas and Symeon Chatzinotas},
  title   = {Attention-Based Blind Adaptive Receive Beamforming
             for Interference Limited {NGSO} Satellite Systems},
  journal = {IEEE Open J. Commun. Soc.},
  volume  = {6},
  pages   = {8946--8964},
  year    = {2025},
  month   = oct,
  doi     = {10.1109/OJCOMS.2025.3622661}
}

@inproceedings{smartut,
  author    = {Almoatssimbillah Saifaldawla and Eva Lagunas and
               Flor Ortiz and Abuzar B. M. Adam and
               Symeon Chatzinotas},
  title     = {{SmartUT}: Receive Beamforming for Spectral
               Coexistence of {NGSO} Satellite Systems},
  booktitle = {Proc. Int. Workshop Smart Antennas},
  pages     = {195--200},
  year      = {2025},
  month     = sep,
  doi       = {10.1109/WSA65299.2025.11202840}
}

@misc{star_constell_gen2,
  author       = {{Federal Communications Commission}},
  title        = {{Space Exploration Holdings, LLC}: Request for orbital deployment and operating authority for the {SpaceX Gen2 NGSO} satellite system},
  howpublished = {FCC 22-91},
  year         = {2022},
  month        = dec
}

@misc{kuiper_eirp,
  author       = {{Kuiper Systems LLC}},
  title        = {Technical Information to Supplement {Schedule S}},
  howpublished = {FCC File No. SAT-LOA-20190704-00057},
  year         = {2019},
  month        = jul
}

@misc{starlink_eirp,
  author       = {{Space Exploration Holdings, LLC}},
  title        = {Technical Information to Supplement {Schedule S}},
  howpublished = {FCC File No. SAT-MOD-20200417-00037},
  year         = {2020},
  month        = apr
}

@misc{star_constell1,
  author       = {{Space Exploration Holdings, LLC}},
  title        = {Application for Modification of Authorization for the {SpaceX NGSO} Satellite System},
  howpublished = {FCC File No. SAT-MOD-20200417-00037},
  year         = {2020},
  month        = apr
}

@misc{kuip_constell1,
  author       = {{Kuiper Systems LLC}},
  title        = {Technical Appendix, Application of {Kuiper Systems LLC} for Authority to Launch and Operate a Non-Geostationary Satellite Orbit System in {Ka}-Band Frequencies},
  howpublished = {FCC File No. SAT-LOA-20190704-00057},
  year         = {2019},
  month        = jul
}

@misc{star_constell2,
  author       = {{Federal Communications Commission}},
  title        = {Order and Authorization and Order on Reconsideration},
  howpublished = {FCC 21-48},
  year         = {2021},
  month        = apr
}

@ARTICLE{kim2025feasibility,
  author={Eunsun Kim and Ian P. Roberts and Jeffrey G. Andrews},
  title={Feasibility Analysis of In-Band Coexistence in Dense {LEO} Satellite Communication Systems},
  journal=IEEE_J_WCOM,
  year={2025},
  volume={24},
  number={2},
  pages={1663--1677},
  month={Feb.},
  doi={10.1109/TWC.2024.3511660}
}

@inproceedings{pyscipopt,
  author    = {Stephen Maher and Matthias Miltenberger and
               Jo{\~a}o Pedro Pedroso and Daniel Rehfeldt and
               Robert Schwarz and Felipe Serrano},
  title     = {{PySCIPOpt}: Mathematical Programming in {Python}
               with the {SCIP} Optimization Suite},
  booktitle = {Proc. Int. Congr. Math. Softw.},
  year      = {2016},
  month     = jul,
  pages     = {301--307}
}

@misc{earthfixed,
  author       = {Xavier Artiga and M{\`a}rius Caus and Ana P{\'e}rez-Neira},
  title        = {Pragmatic {Earth}-Fixed Beam Management for {3GPP NTN} Common Signaling in {LEO} Satellites},
  howpublished = {arXiv preprint arXiv:2512.14368},
  year         = {2025},
  month        = dec
}

@misc{h3,
  author       = {{Uber Technologies, Inc.}},
  title        = {{H3}: Hexagonal Hierarchical Geospatial Indexing System},
  howpublished = {\url{https://h3geo.org/}},
  year         = {2026}
}

\end{document}